\documentclass[doublecol]{epl2}

\usepackage{graphicx}
\usepackage{amsmath}
\usepackage{epstopdf}
\usepackage{amsbsy}
\usepackage{color}
\usepackage{dcolumn}
\usepackage{bm}

\newcommand{\ybco}{YBa$_2$Cu$_3$O$_{6+\delta}$}

\newcommand{\lbco}{La$_{2-x}$Ba$_x$CuO$_4$}

\newcommand{\lnsco}{La$_{2-x-y}$Nd$_y$Sr$_x$CuO$_4$}

\title{Correlation and disorder-enhanced nematic spin response in superconductors with weakly broken rotational symmetry}
\shorttitle{Correlation and disorder-enhanced nematic spin response in superconductors} 

\author{Brian M. Andersen\inst{1} \and Siegfried Graser\inst{2} \and P. J. Hirschfeld\inst{3}}
\shortauthor{B. M. Andersen, S. Graser, and P. J. Hirschfeld}

\institute{
  \inst{1}Niels Bohr Institute, University of Copenhagen, Universitetsparken 5, DK-2100 Copenhagen, Denmark\\
  \inst{2}Theoretical Physics III, Center for Electronic Correlations and Magnetism, Institute
of Physics, University of Augsburg,
D-86135 Augsburg, Germany\\
  \inst{3}Department of Physics, University of Florida, Gainesville, Florida 32611, USA
}
\pacs{73.22.Gk}{Broken symmetry phases}
\pacs{74.20.-z}{Theories and models of superconducting state}
\pacs{74.72.h}{Cuprate superconductors}

\abstract{
Recent experimental and theoretical studies have highlighted the possible role of a electronic nematic liquid in underdoped cuprate superconductors. We calculate, within a model
of $d$-wave superconductor with Hubbard correlations, the spin susceptibility in the case of a small explicitly broken rotational symmetry of the underlying lattice. We then exhibit how the induced  spin response asymmetry is strongly enhanced by correlations as one approaches the instability to stripe order. In the disorder-induced stripe phase, impurities become spin nematogens with a $C_2$ symmetric impurity resonance state, and the disorder-averaged spin susceptibility remains only $C_2$ symmetric at low energies, similar to recent data from neutron scattering experiments on underdoped YBCO.
}

\begin{document}

\maketitle

\section{Introduction}
Incommensurate one dimensional composite spin and charge density waves,
often called ``stripes''\cite{pnas,jan,antonio_rev,brom_rev,kiv_rmp,oles_rev},
have been observed and play an important role in discussions of the
underdoped cuprates and other systems.  Such states were  predicted theoretically in
the context of mean-field studies of Hubbard models\cite{za89,poil89,schulz89,machida89},
and later observed in neutron scattering experiments in \lbco\ and \lnsco \cite{jt95,jt96}.
Stripes break the discrete translation and rotation symmetries of the
 CuO$_2$ planes.  Rotational symmetry is also broken in so-called liquid crystal analogs called ``electronic nematic'' states,  but these preserve translational symmetry and  may occur as the initial instability of
a paramagnetic state  before an ordered state of charge, spin, or combined spin and charge order
is reached\cite{kiv_rmp,KFE98,vojta_review,fradkin,vojta}.

While stripe-like ground states were originally thought to be a very special
feature of the 214 compounds, this view has changed in recent years, in particular
with the discovery of broken $C_4$ symmetry in the spin response of highly underdoped samples of \ybco \cite{hinkov08a},
and, more recently, a subtle charge order in the same system\cite{MHJulien}. Signatures of nematic order have also been recently reported in transport and tunneling measurements\cite{ando,daou,lawler}. In all these cases, however,
the question of $C_4\rightarrow C_2$ symmetry breaking is muddied by the fact that the crystal is
not tetragonal, since for example in \ybco\  the CuO chains give a well-defined anisotropy in the untwinned samples on which
the experiments were performed, such that the system is formally orthorhombic.  Nevertheless, the evolution
from optimally to highly underdoped samples, which is accompanied by a dramatic enhancement of the anisotropy
of the responses, is quite striking, and leads to the common assumption that these highly correlated underdoped materials
display a strongly enhanced ``nematic susceptibility", i.e. a tendency to create nematic order which is
driven by the very small symmetry-breaking field provided by the $x-y$ anisotropy in the band-structure.  However,
these ideas have rarely been cast in a concrete microscopic model allowing
a direct study of how disorder and local electronic correlations which drive the Mott insulating state
 influence nematicity. Previous studies have mainly focussed on the high-energy spin fluctuations of the RPA susceptibility of a homogeneous $d$-wave
 superconductor with an anisotropic band-structure\cite{li,kao,eremin,schnyder,yamase}, or utilized phenomenological Ginzburg-Landau approaches\cite{vojta,kaul,sun}. More recently, the nematic response has also been studied within the two-dimensional Hubbard model with slight $x-y$ hopping asymmetry using strong-coupling cluster methods, and found to be significantly enhanced by interactions at low temperatures in the underdoped pseudogap regime\cite{okamoto,maier}.

The salient features of the inelastic neutron scattering experiments
on strongly underdoped untwinned YBCO samples which should be reproduced by
a reasonably complete theoretical analysis are as follows:
\begin{enumerate}
  \item The  neutron intensity near ($\pi,\pi$) evolves from a pattern of four symmetrically placed incommensurate peaks at high energies which merge at ($\pi,\pi$) at the spin resonance energy $\Omega_0$, to a pattern with $C_2$ symmetry with two incommensurate
peaks at quite low energies $\omega \ll \Omega_0$.
  \item The details of this pattern appear to be important: the intensity in the nematic case has a saddle point form in $q$ space, with a maximum at ($\pi,\pi$) for cuts along $b$ and a minimum there for cuts along $a$.
  \item The strength of the $C_4$ symmetry breaking increases as one underdopes.
\end{enumerate}
Some phenomenologies have been rather successful in accounting for some of these features, but to our knowledge there is no microscopic approach which has thus far successfully accounted for all of them.
In this work we present a simple version of
such a theory, based on our earlier work exploring the effect of Hubbard correlations on the $d$-wave superconducting
state, to which we now add a small symmetry-breaking field to simulate e.g. the effect of the chains in \ybco.
We show that the nematicity in the spin response is enhanced by correlations and decreasing
temperature; what is perhaps more surprising is our finding that the phenomenon can be enhanced further by pairing and
disorder. We find that disorder is crucial to explain the saddle point
    structure of the inelastic scattering intensity in $q$-space.
    In addition, we present local investigations which exhibit the explicit formation
of nematogens, nematically driven impurity states, which have not to our knowledge been observed in
the cuprates, although recently reported in the stripe-ordered phase of Fe-based systems\cite{Davis,Curro,Hanaguri}.
Indications of an enhanced nematic susceptibility above the
magnetic transition have also been observed in these systems\cite{pnictide_nematic}.

\section{Model}

The Hamiltonian is given by
\begin{eqnarray}\nonumber\label{Hamiltonian}
\hat{H} &\!\!\!\!=\!\!\!\!& -\!\!\sum_{ij\sigma}t_{ij}\hat{c}_{i\sigma}^{\dagger}\hat{c}_{j\sigma}\!\!+\!\!\sum_{i\sigma}(V_{i}\!\!-\!\!\mu)\hat{n}_{i\sigma} +U\sum_{i\sigma}\frac{n_{i}-\sigma m_{i}}{2}\hat{n}_{i\sigma}\\
 &+&\sum_{i\delta}\left( \Delta_{\delta i}\hat{c}_{i\uparrow}^{\dagger}\hat{c}_{i+\delta\downarrow}^{\dagger}+\Delta_{\delta i}^{*}\hat{c}_{i+\delta\downarrow}\hat{c}_{i\uparrow}\right),
\end{eqnarray}
where $\hat{c}_{i\sigma}^\dagger$ creates an electron on site $i$
with spin $\sigma$, and $t_{ij}=\{t_x,t_y,t'\}$ denote the hopping integrals to the two nearest neighbors. In Eq.(\ref{Hamiltonian}), $n_i$ and $m_i$ refer to the charge density and magnetization, respectively, $V_i$ is an impurity potential from a
set of $N$ point-like scatterers, $\mu$ is the chemical potential
and $\Delta_{ij}$ is the $d$-wave pairing potential between sites
$i$ and $j$. The amplitude of $\Delta_{ij}$ is set by the superconducting
coupling constant $g$ and will exhibit a slight anisotropy inherited from a finite $\delta_0=(t_y-t_x)$. Below we fix the parameters
$t'=-0.35t$, adjust $\mu$ to give a hole doping $x=1-n\simeq 10\%$, and $g=0.6$ leading to realistic pairing amplitudes $\Delta=0.10t$. The hopping asymmetry $\delta_0=0.05$ is fixed for all results presented here, and we use units where $t=t_y=1.0$. We have solved
Eq.(\ref{Hamiltonian}) self-consistently on unrestricted  $N\times N$ lattices by diagonalizing the
associated Bogoliubov-de Gennes (BdG) equations at $T=0.01t$.\cite{JWHarter:2006}

The model given by Eq.(\ref{Hamiltonian}) has been used
extensively in the literature to study the competition between bulk superconducting and magnetic phases, and
field-induced magnetization\cite{allHamiltonian}. It has also been used to
study moment formation around nonmagnetic impurities
in correlated $d$-wave superconductors\cite{alloul09}. In
the case of many impurities, Eq.(\ref{Hamiltonian}) was used to model static disorder-induced magnetic droplets\cite{alvarez,BMAndersen:2006,atkinson,andersen07,andersen08}, and explain how these may increase in volume fraction when moving to lower doping levels and eventually form a  quasi-long-range ordered magnetic stripe phase.
More recently, Eq.(\ref{Hamiltonian}) extended to the vortex state was used to obtain a semi-quantitative description of the temperature dependence of the elastic neutron response in underdoped LSCO\cite{schmid2,AndersenJPCS}.

The transverse bare spin susceptibility $\chi_0^{xx}(\vec{r}_{i},\vec{r}_{j},\omega)=-i\int_{0}^{\infty}dt\, e^{i\omega t}\left\langle \left[\sigma_{i}^{x}(t),\sigma_{j}^{x}(0)\right]\right\rangle$, can be expressed in terms of the BdG eigenvalues $E_n$ and eigenvectors $u_n, v_n$ as
\begin{eqnarray}\nonumber
\chi_0^{xx}(\vec{r}_{i},\vec{r}_{j},\omega) & = & \sum_{m,n} f(u,v)\frac{f(E_{m})+f(E_{n})-1}{\omega+E_{m}+E_{n}+i\Gamma}\\
 & + & \sum_{m,n} g(u,v)\frac{f(E_{m})+f(E_{n})-1}{\omega-E_{m}-E_{n}+i\Gamma},
\end{eqnarray}
\begin{eqnarray}
f(u,v) & = & u_{m,i}^{*}v_{n,i}^{*}\left(u_{m,j}v_{n,j}-u_{n,j}v_{m,j}\right),\\
g(u,v) & = & v_{m,i}u_{n,i}\left(u_{m,j}^{*}v_{n,j}^{*}-u_{n,j}^{*}v_{m,j}^{*}\right).
\end{eqnarray}

Including the electronic interactions within RPA we find for the full susceptibility
\begin{equation}
\chi^{xx}(\vec{r}_i,\vec{r}_j,\omega)\!=\!\sum_{\vec{r}_l}\left[1-U\chi^{xx}_{0}(\omega)\right]_{\vec{r}_i,\vec{r}_l}^{-1}\chi^{xx}_{0}(\vec{r}_l,\vec{r}_j,\omega).
\end{equation}
Fourier transforming with respect to the relative coordinate $\vec{r}=\vec{r}_i-\vec{r}_j$ defines the spatially resolved momentum-dependent
susceptibility $\chi(\vec{q},\vec{R},\omega)=\sum_{\vec{r}}e^{i\vec{q}\cdot\vec{r}}\chi(\vec{R},\vec{r},\omega)$. Averaging over the center of mass coordinate $\vec{R}=(\vec{r}_i+\vec{r}_j)/2$, this expression gives the susceptibility $\chi({\bf q},\omega)$ relevant for comparison with neutron measurements.

\section{Results}
\begin{figure}[h!]
\begin{minipage}{.49\columnwidth}
\includegraphics[clip=true,width=0.98\columnwidth]{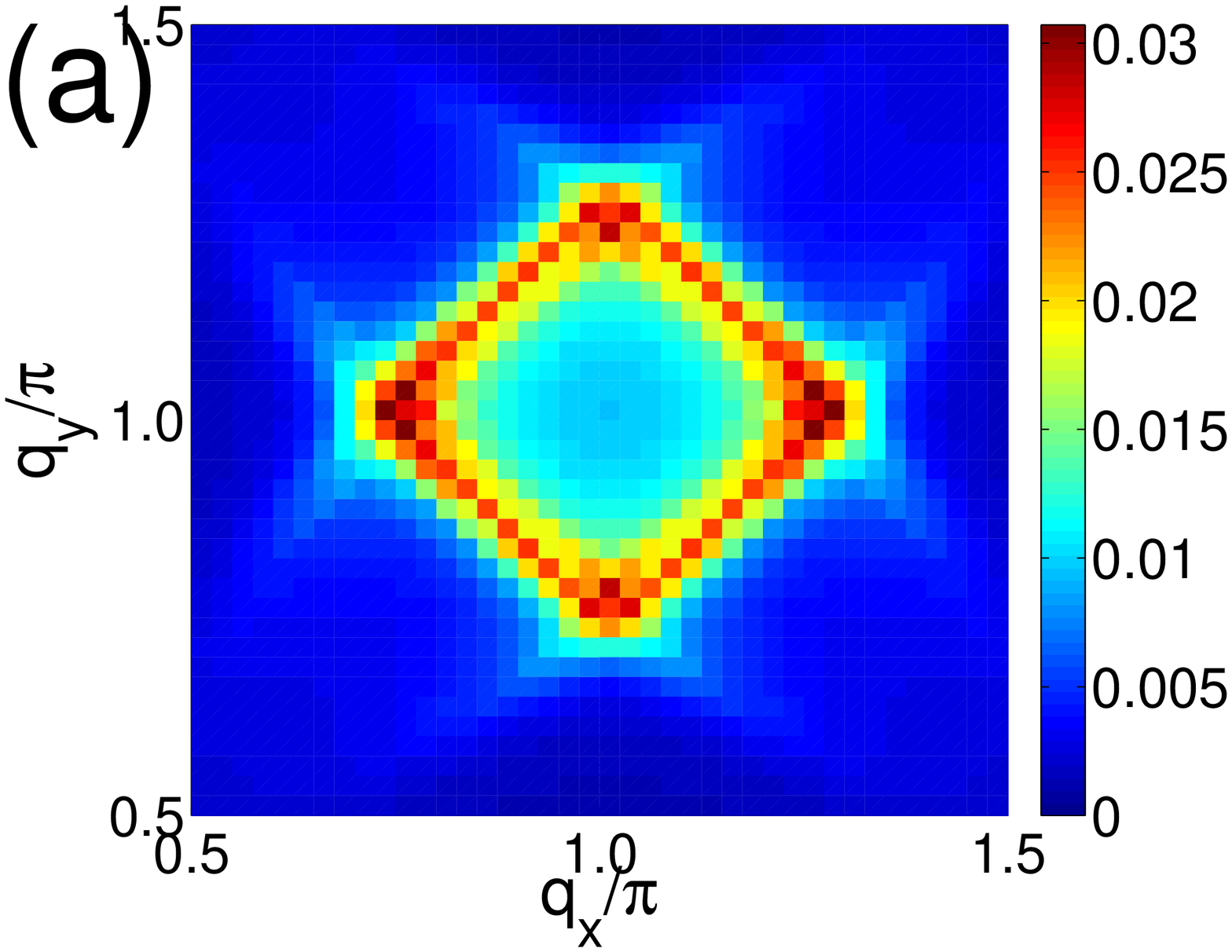}
\end{minipage}
\begin{minipage}{.49\columnwidth}
\includegraphics[clip=true,width=0.98\columnwidth]{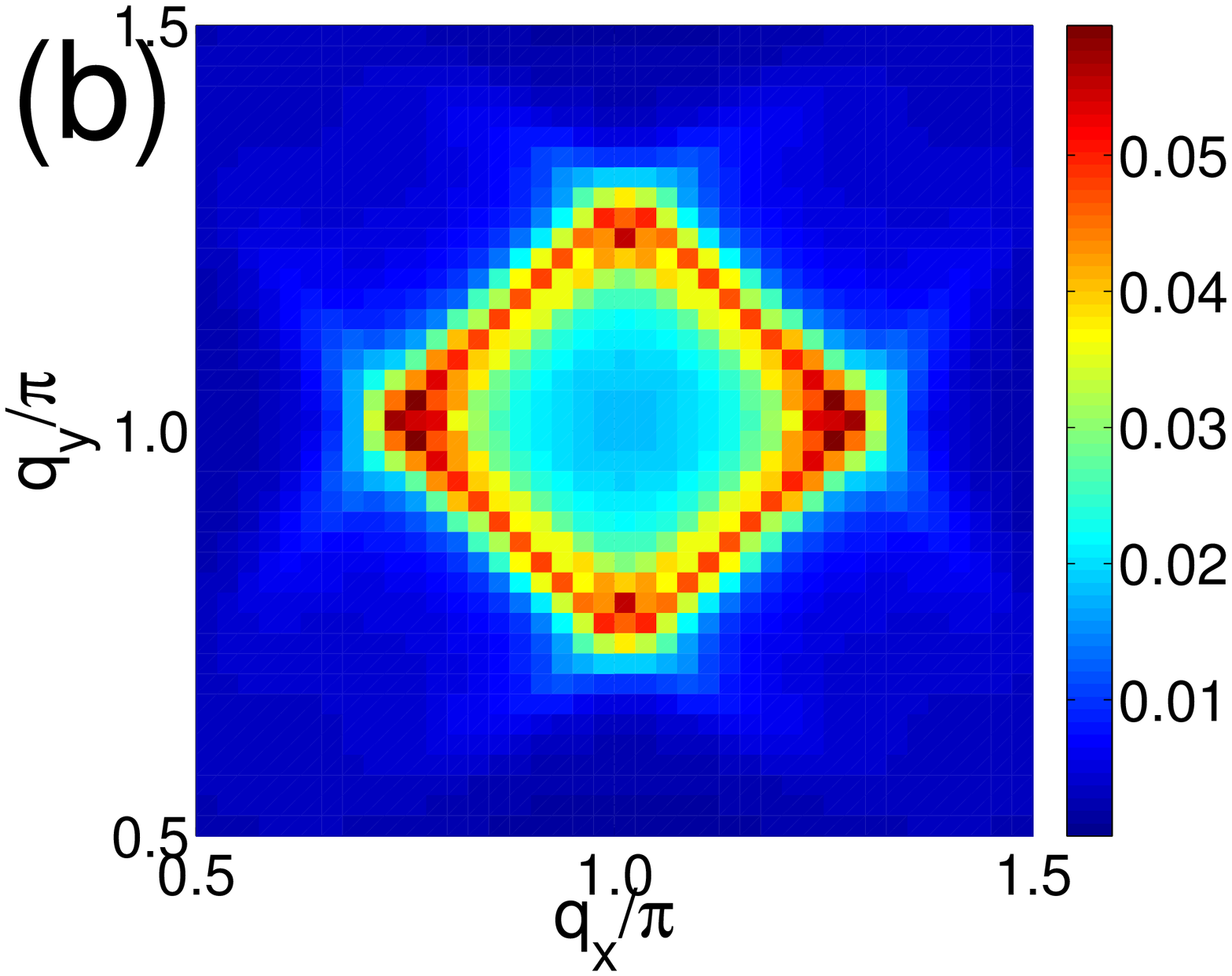}
\end{minipage}
\begin{minipage}{.49\columnwidth}
\includegraphics[clip=true,width=0.98\columnwidth]{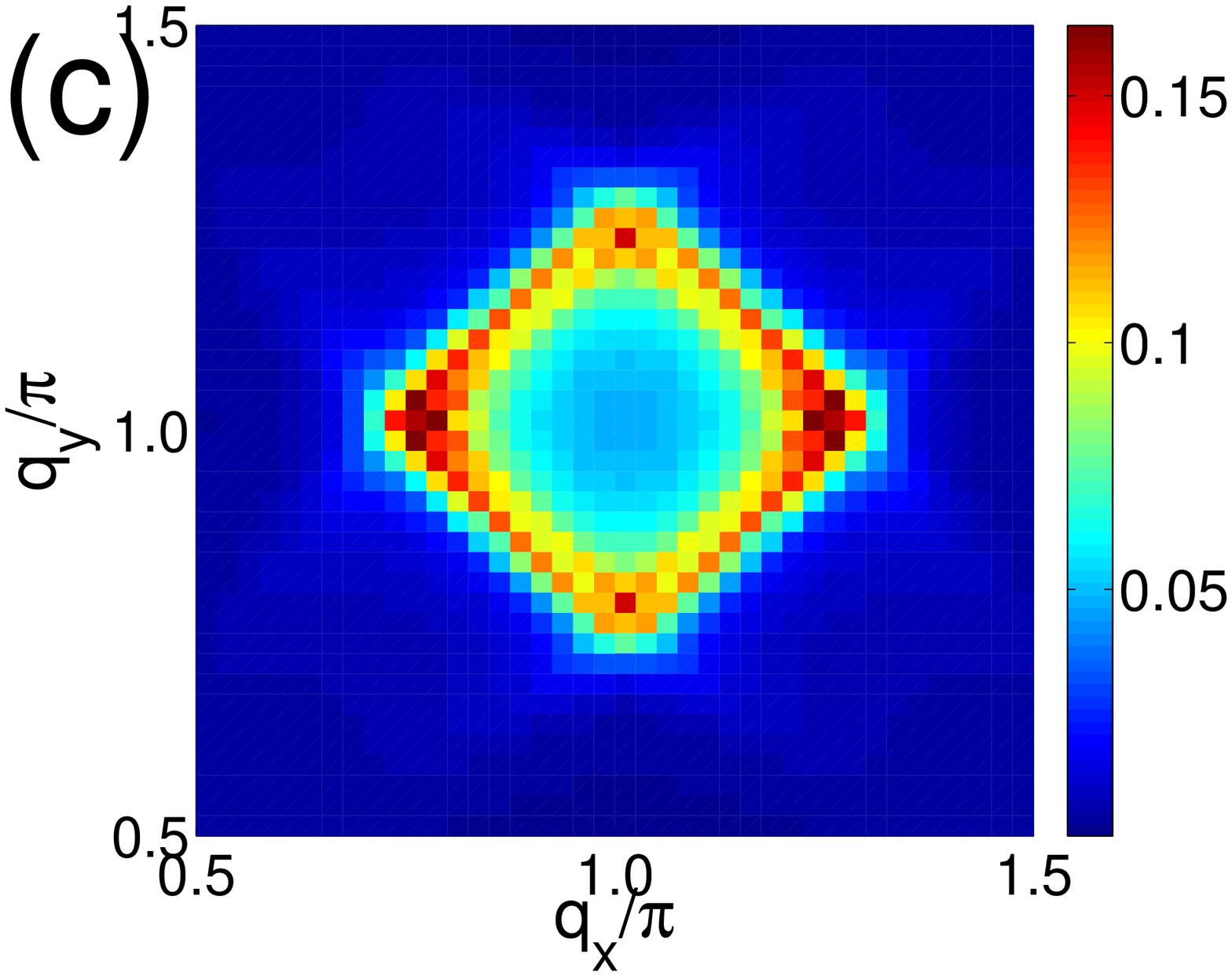}
\end{minipage}
\begin{minipage}{.49\columnwidth}
\includegraphics[clip=true,width=0.98\columnwidth]{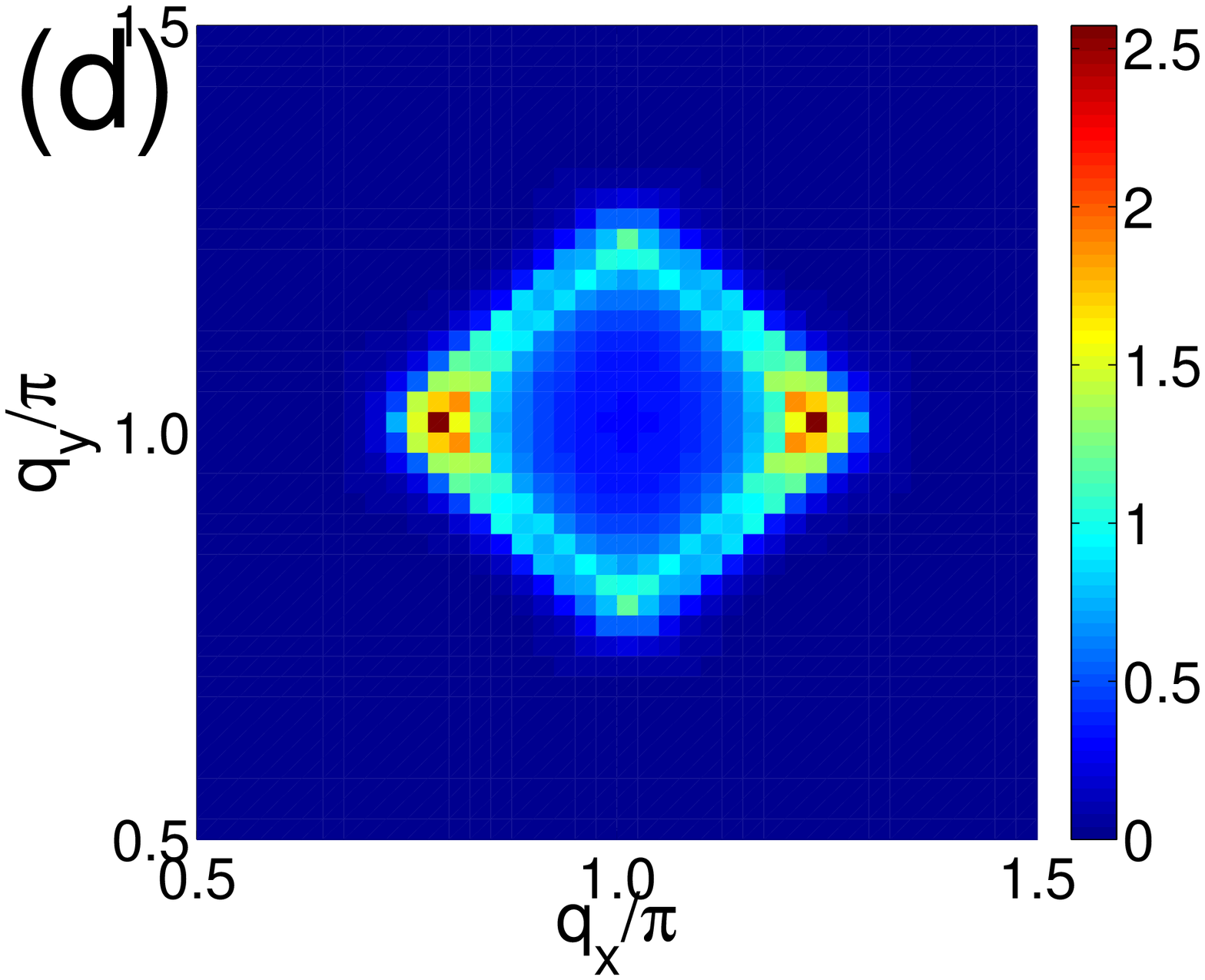}
\end{minipage}
\caption{(Color online) Constant energy cuts of the susceptibility $\chi''(q,\omega)$ at low energy $\omega/t=0.0$ for $\delta_0=0.05$, $N=80$, and $U/t=2.0$ (a), $U/t=2.2$ (b), $U/t=2.4$ (c), and $U/t=2.6$ (d).} \label{cutasym005vsU}
\end{figure}

\begin{figure}[h!]
\includegraphics[clip=true,width=0.98\columnwidth]{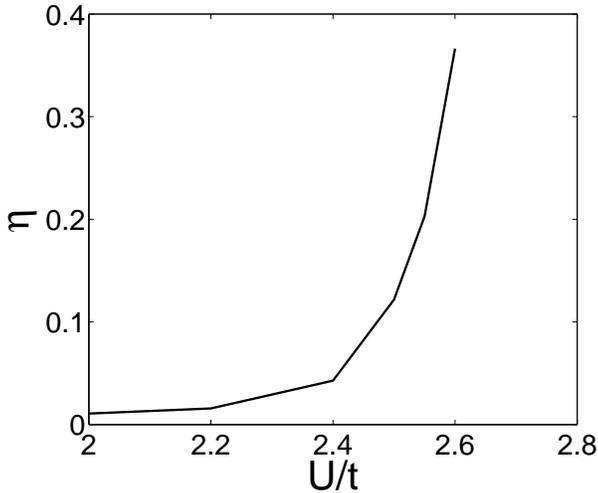}
\caption{Nematic spin response $\eta$ versus $U$ for $\delta_0 =0.05$ in the low-$U$ homogeneous $d$-wave superconducting phase.} \label{NvsU}
\end{figure}

\begin{figure}[]
\begin{minipage}{.49\columnwidth}
\includegraphics[clip=true,width=0.98\columnwidth]{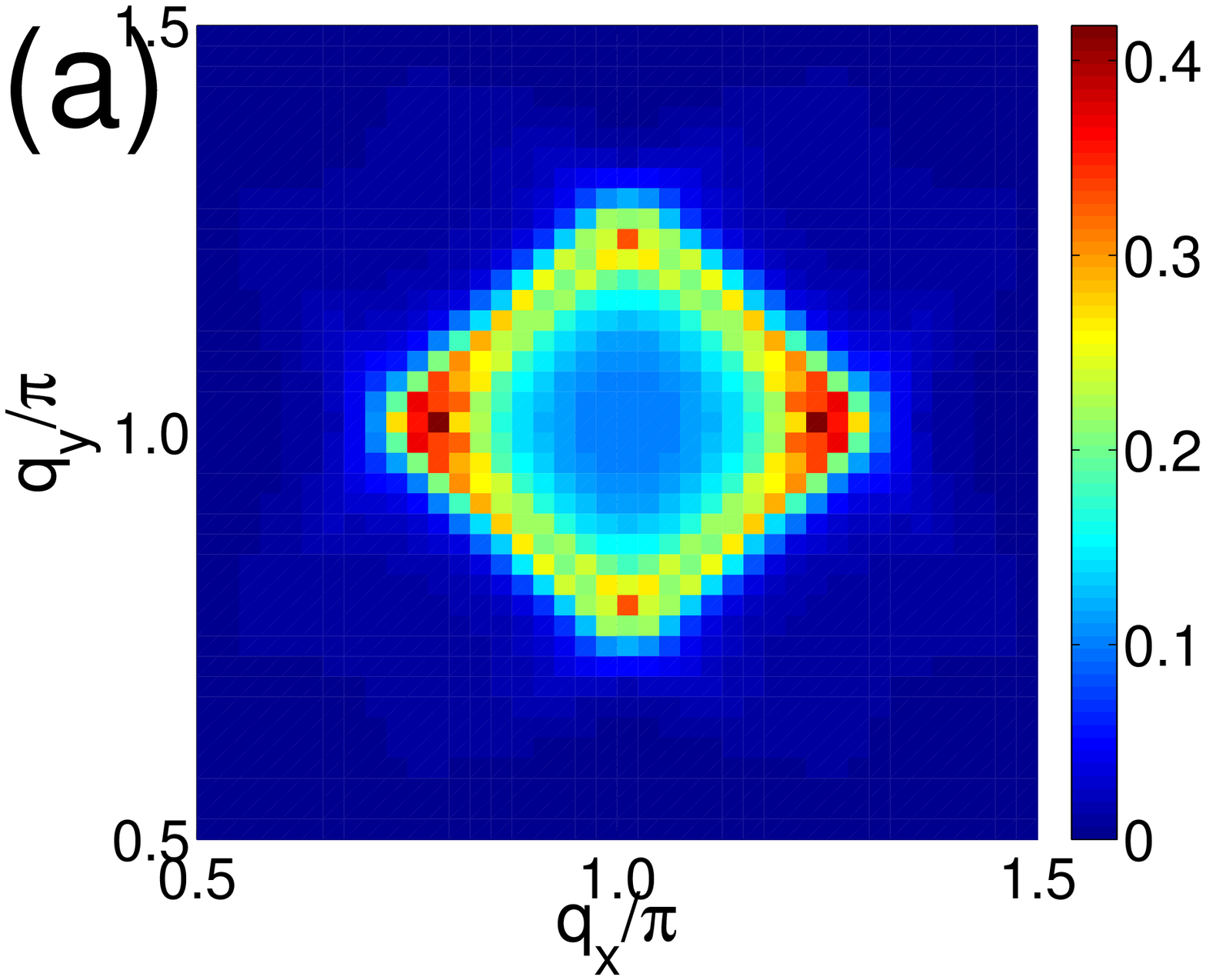}
\end{minipage}
\begin{minipage}{.49\columnwidth}
\includegraphics[clip=true,width=0.98\columnwidth]{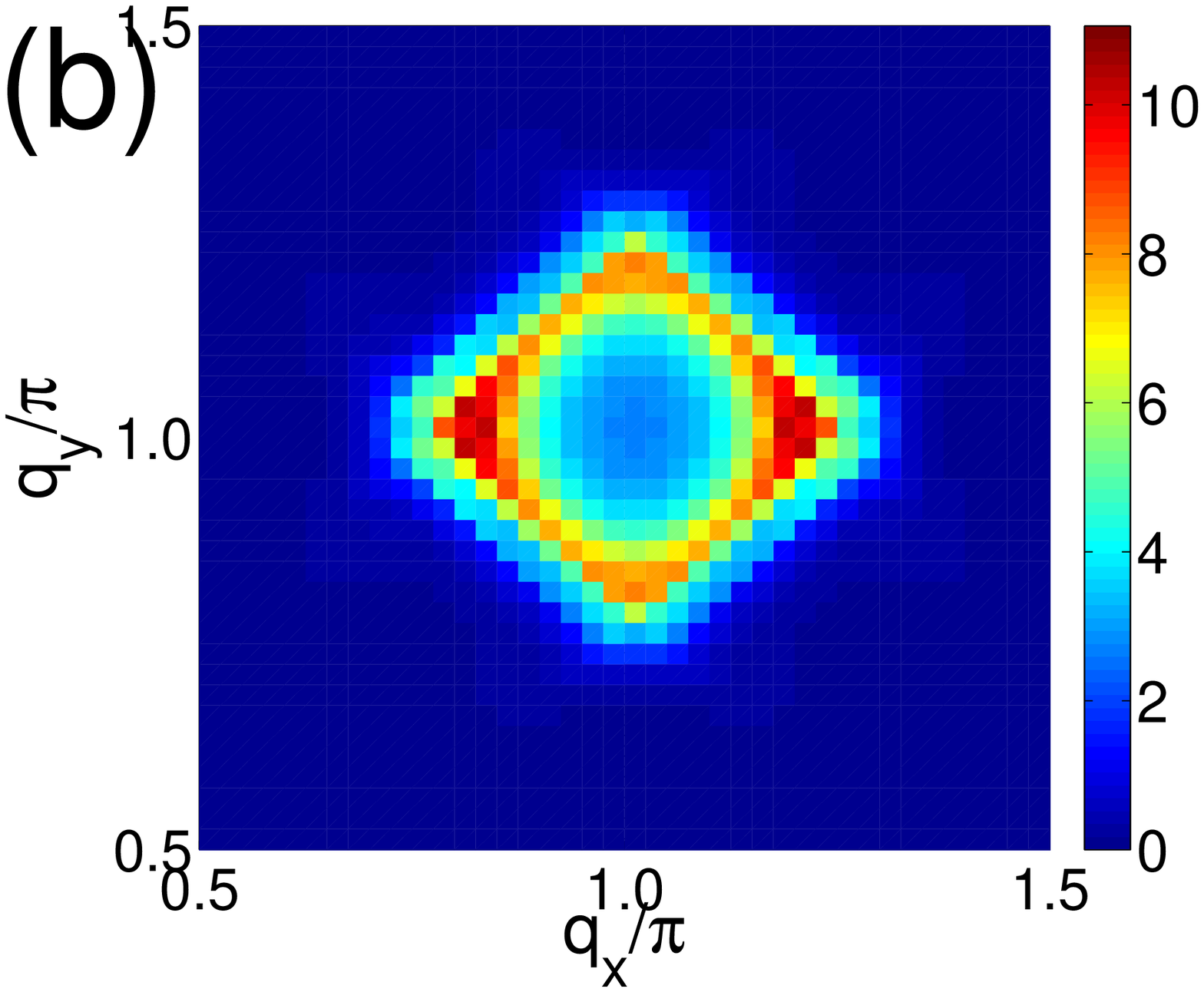}
\end{minipage}
\begin{minipage}{.49\columnwidth}
\includegraphics[clip=true,width=0.98\columnwidth]{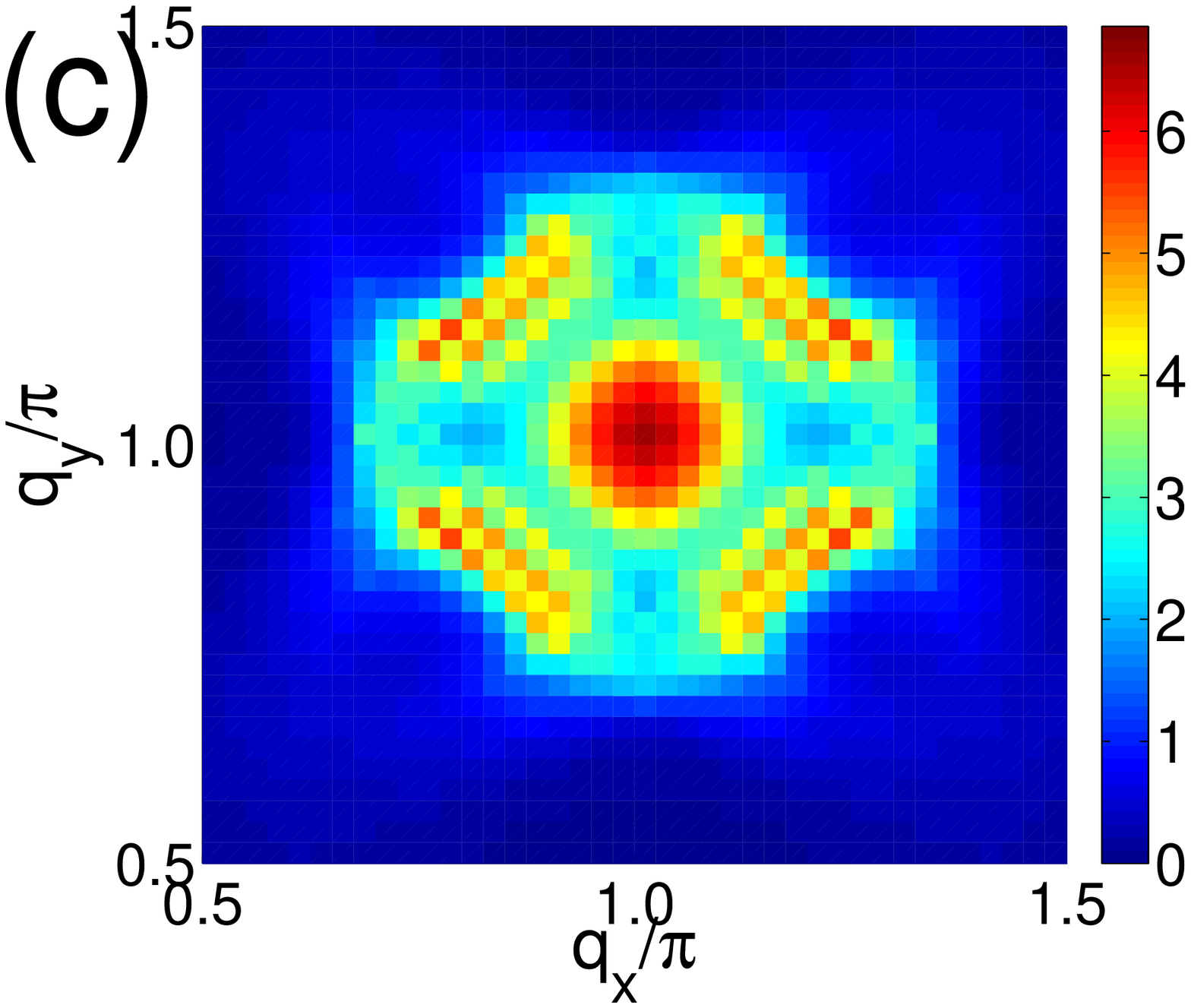}
\end{minipage}
\begin{minipage}{.49\columnwidth}
\includegraphics[clip=true,width=0.98\columnwidth]{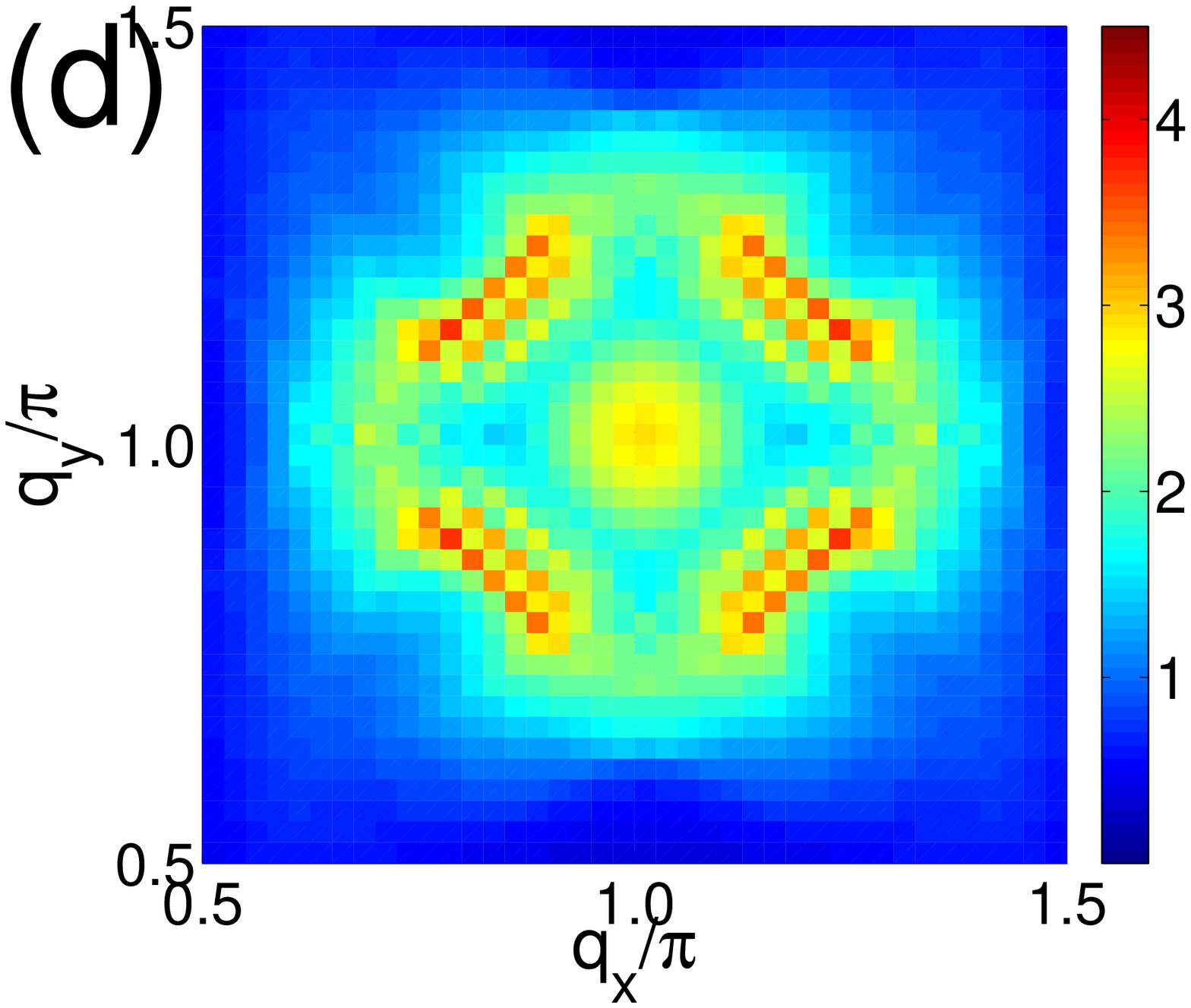}
\end{minipage}
\caption{(Color online) Constant energy cuts of the susceptibility $\chi''(q,\omega)$ with $U/t=2.5$, $\delta_0=0.05$, $N=80$, and $\omega/t=0.0t$ (a), $\omega/t=0.20t$ (b), $\omega/t=0.40t$ (c), and $\omega/t=0.60t$ (d).} \label{cutU25asym005}
\end{figure}

Figure \ref{cutasym005vsU} shows the susceptibility $\chi''(q,\omega)=\mbox{Im} \chi({\bf q},\omega)$ at low energy $\omega/t=0.0$ versus $q_x$ and $q_y$ for different $U$ in the homogeneous superconducting phase. As seen, the apparent asymmetry in the neutron response is clearly enhanced as the correlations are enhanced. Quantitatively, we can define a nematic spin-response "order parameter" as
\begin{equation}
\eta=\frac{\mbox{max}\{ \chi''(q_x,\pi,\omega_0)\} - \mbox{max}\{ \chi''(\pi,q_y,\omega_0)\}}{\mbox{max}\{ \chi''(q_x,\pi,\omega_0)\} + \mbox{max}\{ \chi''(\pi,q_y,\omega_0)\}}.
\end{equation}
Figure \ref{NvsU} shows the dramatic increase of $\eta$ as $U$ approaches the stripe instability from below for fixed hopping asymmetry $\delta_0 = 0.05$. In cuprates with weak orthorhombicity, it is therefore natural to expect a significant enhancement of the nematic response upon lowering the doping level from the optimally doped regime\cite{vojta,okamoto,maier}.

Experimentally, it is known that the observed $x-y$ asymmetry disappears as one increases the energy $\omega$ which seems natural in light of the presumably very small energy scale associated with the nematic aligning field. Here, the same result is obtained as seen from Fig. \ref{cutU25asym005} where the panels show constant-energy cuts of $\chi''(q,\omega)$ at varying $\omega$ with fixed $U=2.5t$ and $\delta_0=0.05$.

The discussion above, and the associated results presented in Figs.\ref{cutasym005vsU}-\ref{cutU25asym005}, were all based on a ground state consisting of a homogeneous $d$-wave superconductor. Within the present mean-field formalism this is the relevant regime for clean systems with $U<U_{c2}\simeq 2.6t$. In the opposite regime, where $U>U_{c2}$, the clean phase is an ordered smectic stripe phase which disorder may turn into a nematic as discussed e.g. in Refs. \cite{andersen10,robertson,delmaestro}. We do not expect a small explicit symmetry breaking term to qualitatively alter this scenario. Instead, we focus on the weak-$U$ limit where disorder may induce a stripe magnetic phase as shown in Refs. \cite{alvarez,atkinson,andersen07,andersen08}, and study the evolution of the nematic spin response to disorder. For this purpose we need to investigate how impurities respond to a finite $\delta_0$.

\begin{figure}[]
\begin{minipage}{.49\columnwidth}
\includegraphics[clip=true,width=0.98\columnwidth]{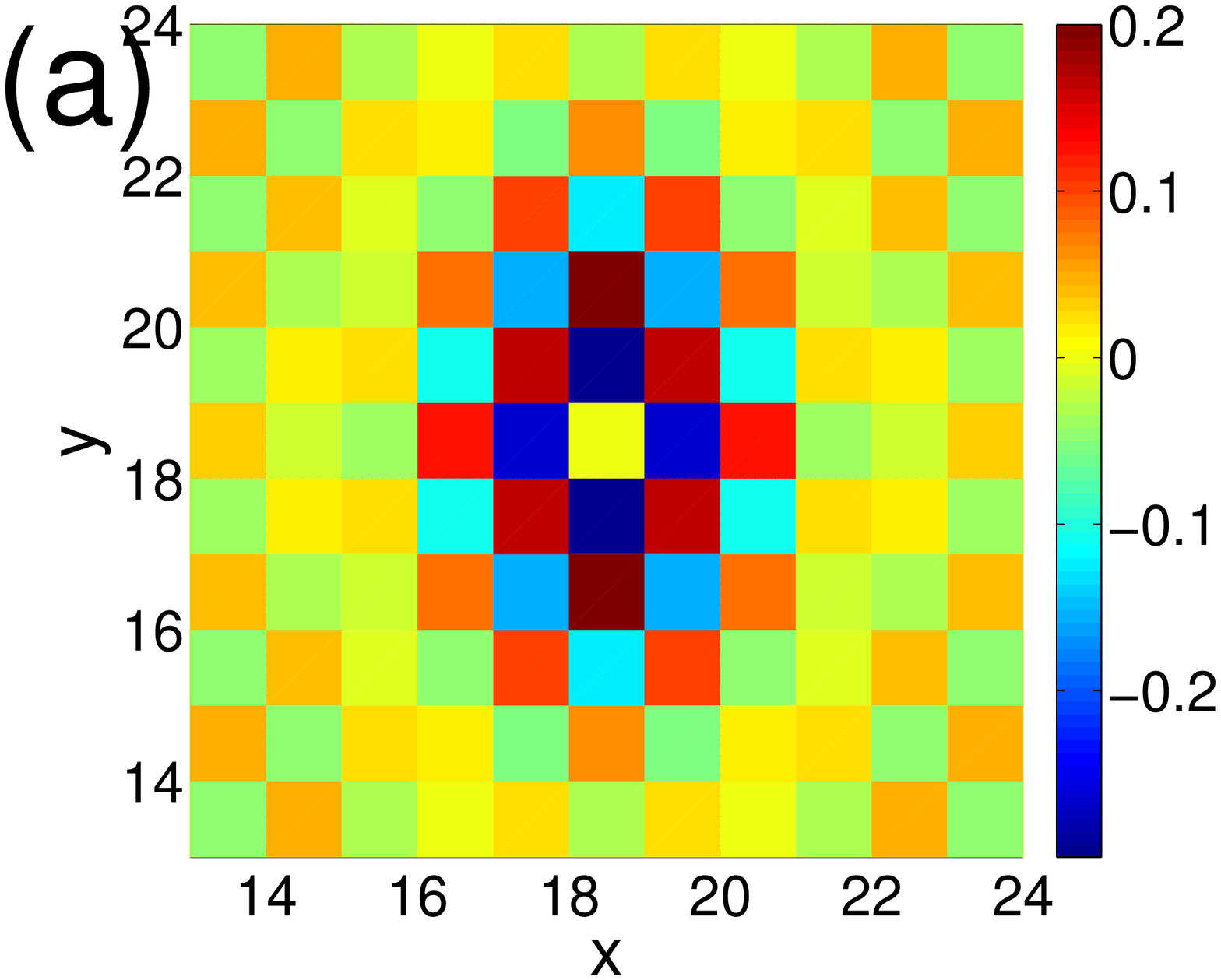}
\end{minipage}
\begin{minipage}{.49\columnwidth}
\includegraphics[clip=true,width=0.98\columnwidth]{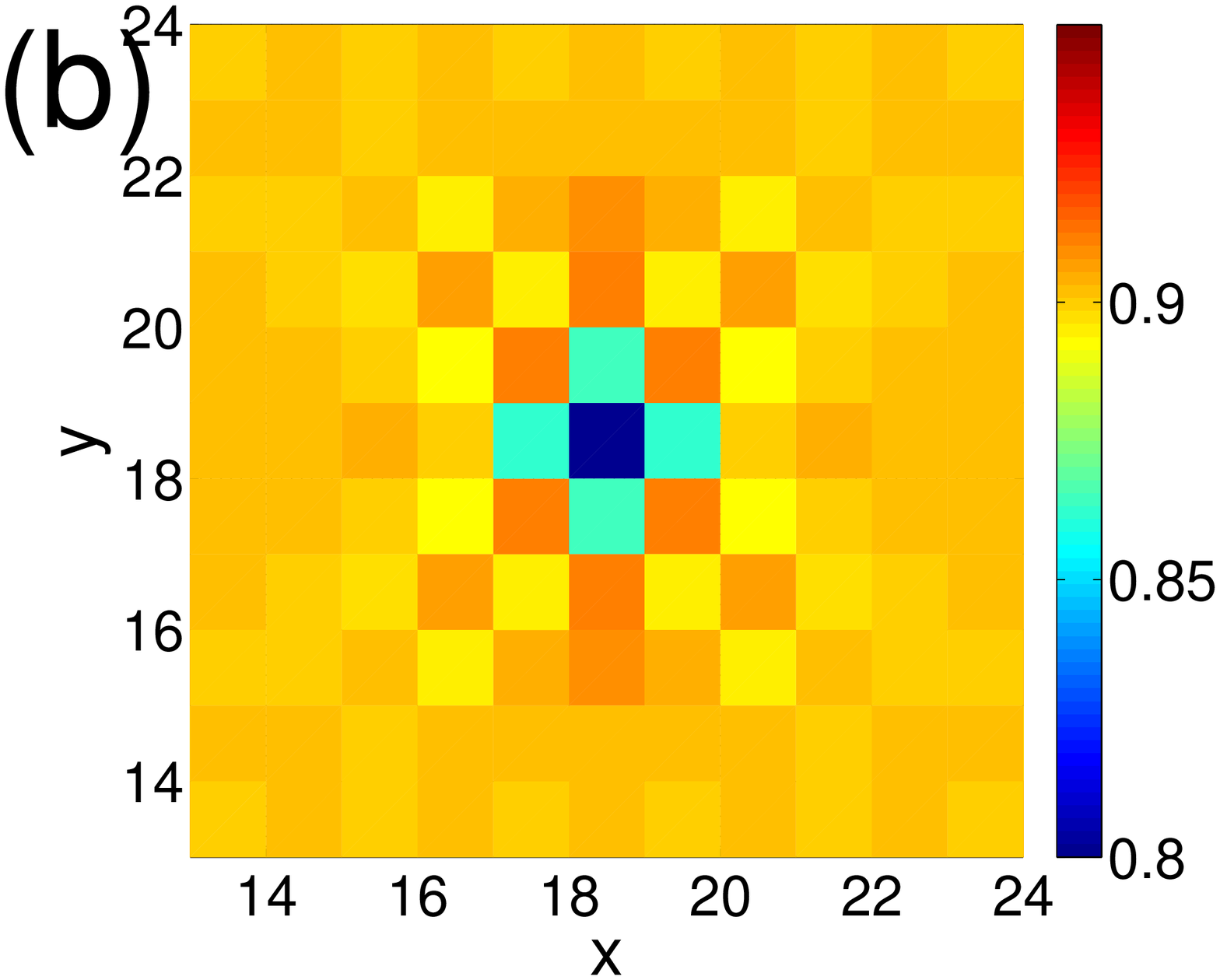}
\end{minipage}
\begin{minipage}{.49\columnwidth}
\includegraphics[clip=true,width=0.98\columnwidth]{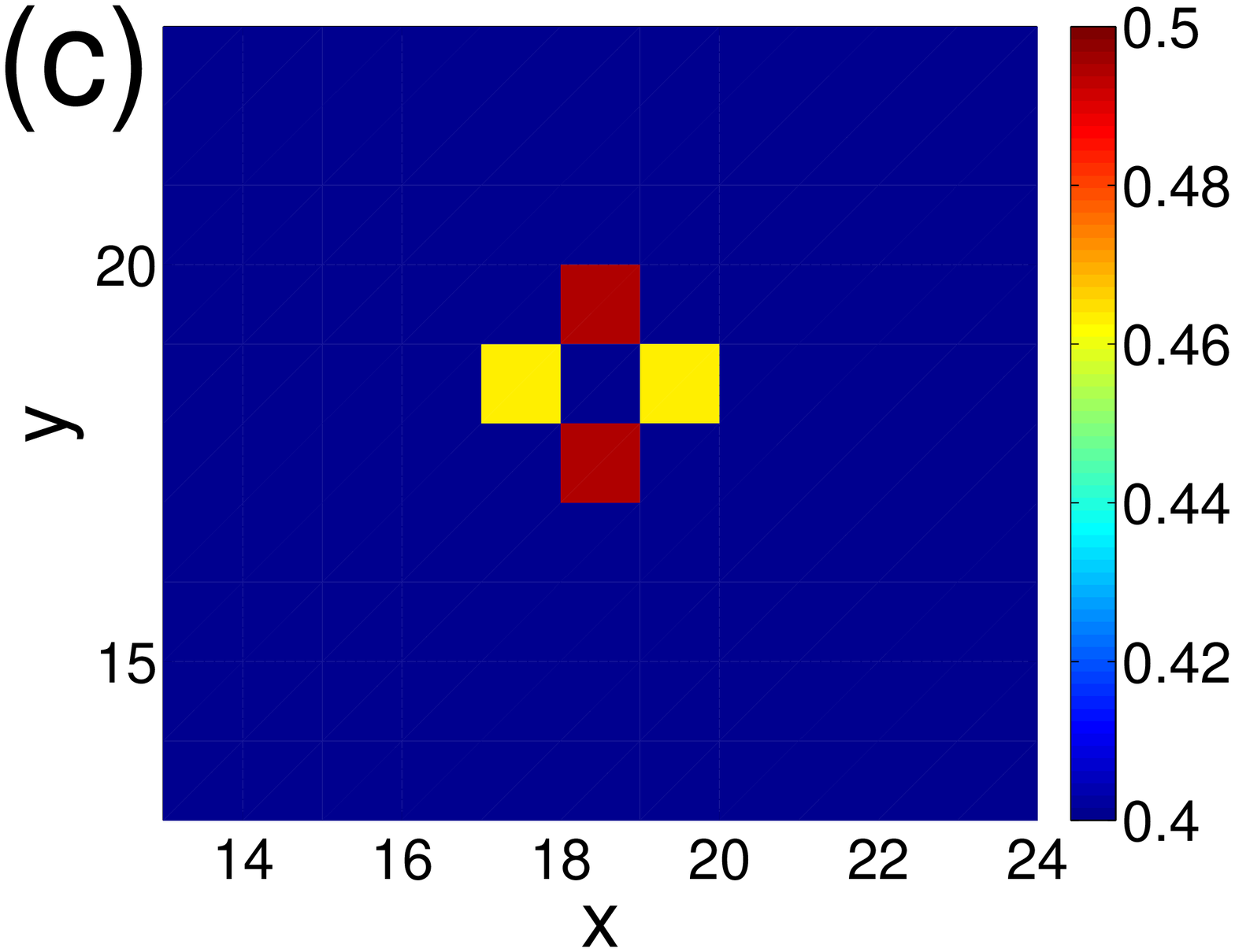}
\end{minipage}
\begin{minipage}{.49\columnwidth}
\includegraphics[clip=true,width=0.98\columnwidth]{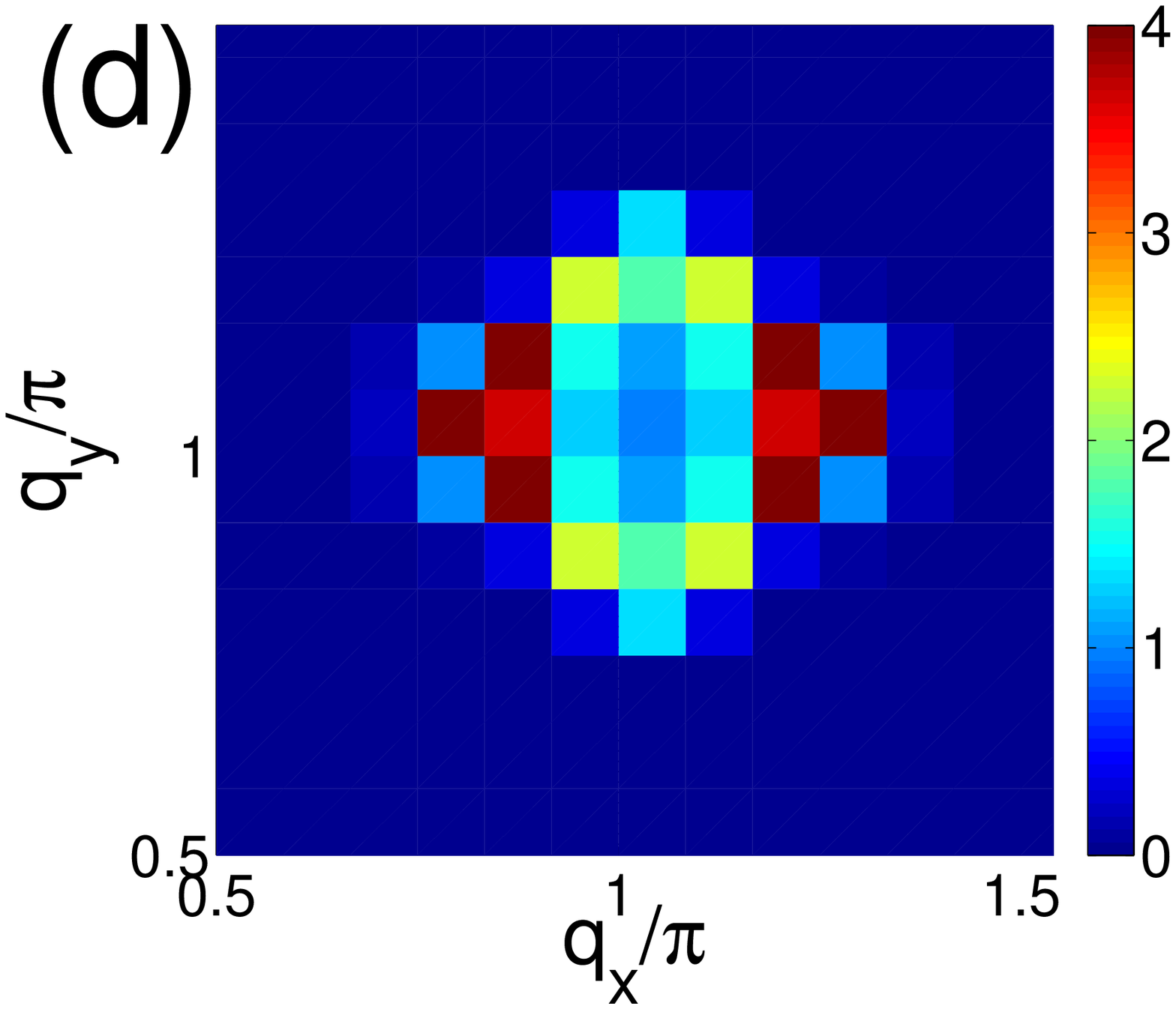}
\end{minipage}
\begin{minipage}{.49\columnwidth}
\includegraphics[clip=true,width=0.98\columnwidth]{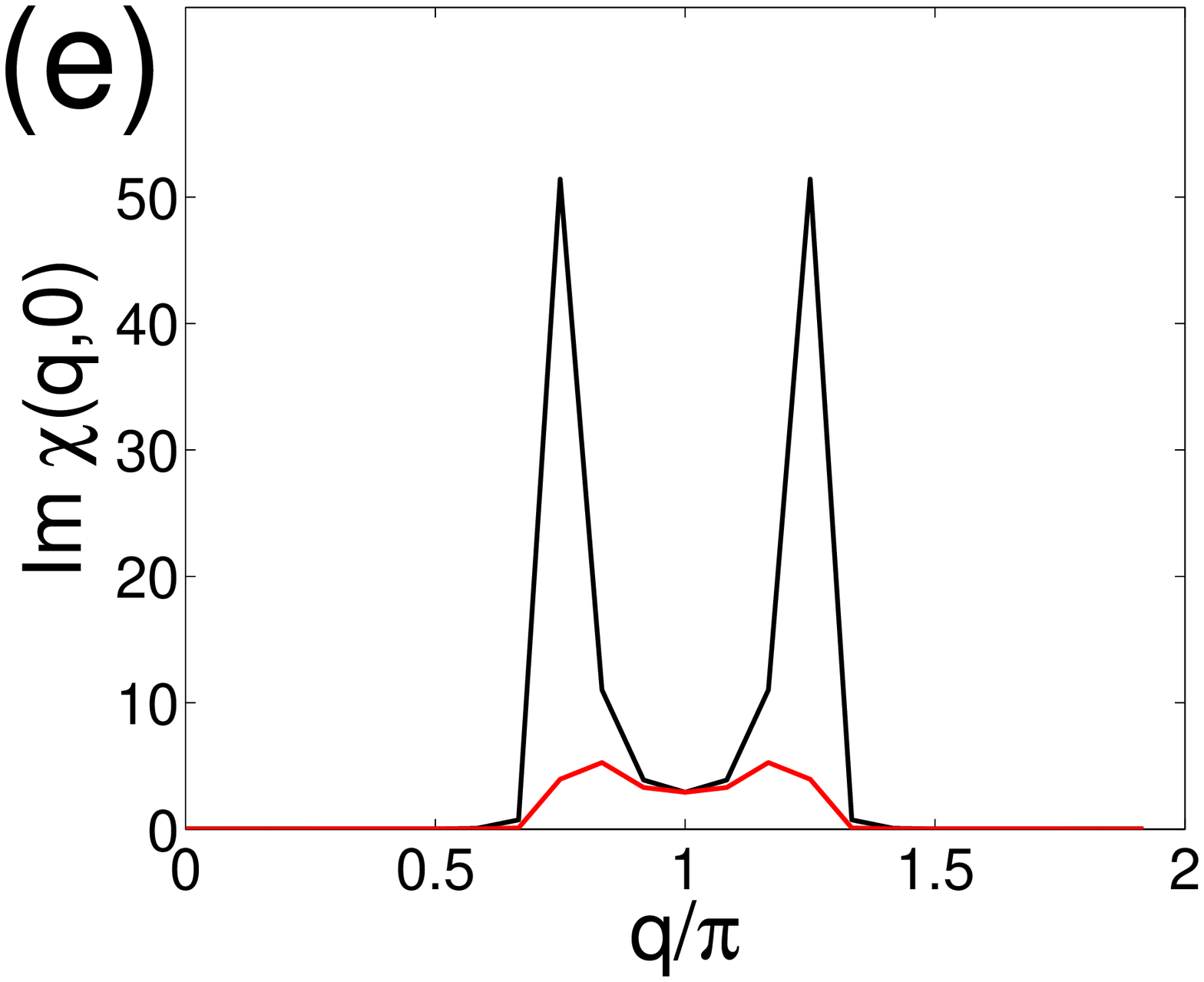}
\end{minipage}
\begin{minipage}{.49\columnwidth}
\includegraphics[clip=true,width=0.98\columnwidth]{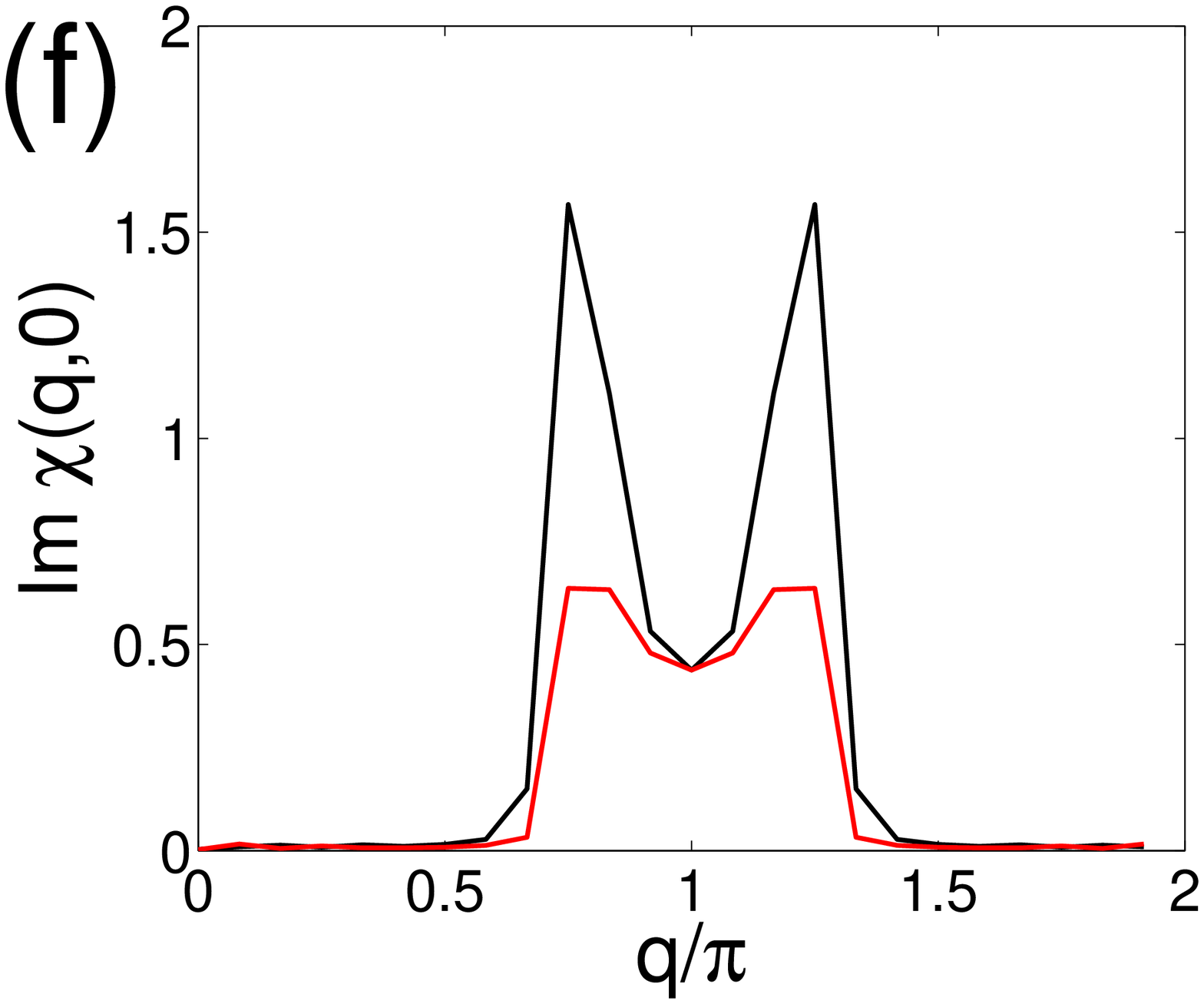}
\end{minipage}
\caption{(Color online) Magnetization (a), charge density (b), and local density of states (c) plotted in real-space near a single strong scatterer ($V_{imp}=100t$) with hopping asymmetry $\delta_0=0.05$ and $U/t=2.5$. In (a-c) we have used $N=35$, but show only the center $11\times 11$ sites near the impurity for clarity. In (c) the LDOS is shown at $\omega=-0.05t$ which is the energy of the impurity resonance for the present choice of parameters. (d) Constant energy cut of the susceptibility $\chi''(q,\omega)$ with $U/t=2.5$ and $\omega/t=0.0t$ for the case of disorder-induced nematogens corresponding to (a). (e) shows $q_x$ ($q_y$) cuts of $\chi''(q,\omega)$ along $q_y=\pi$ [black line] ($q_x=\pi$ [red line]) for the same case as shown in panel (d). (f) same as (e) but in the homogeneous case similar to Fig. \ref{cutU25asym005}(a).} \label{nematodes}
\end{figure}

Figure \ref{nematodes}(a) shows the $C_2$ symmetric pattern of local magnetic order induced around such an impurity,
which we will refer to as a ``spin nematogen", while in  Figure \ref{nematodes}(b) we  see that a significant anisotropy is absent in the charge sector, as may be expected in weak-coupling theories of this type.
 Similar to the bulk results in Fig. \ref{cutasym005vsU}, the impurity-induced magnetization becomes increasingly asymmetric as one approaches $U_{c2}$ from below. The question arises how to detect these spin nematogens in the cuprates. NMR is clearly sensitive to the local distribution of spins, and the lineshape in other disordered, correlated materials has been analyzed
 in terms of postulated structures of this type\cite{Curro}. In addition, one might expect a signature in the local density of states around an impurity; the well known fourfold-symmetric impurity resonance  associated with a potential scatterer in a $d$-wave superconductors has been detected by STM\cite{pan} in near-optimally doped samples. In the presence of $x-y$ anisotropy, this resonance acquires within our theory a similar anisotropy as shown in Fig. \ref{nematodes}(c) which should be observable near strong scatterers in the underdoped regime.

The spin nematogens around an impurity arise because of a "freezing" of gapped incommensurate spin fluctuations present in the clean system\cite{AndersenJPCS,andersen10,suchaneck}. In agreement with the $x-y$ anisotropy shown in Fig. \ref{nematodes}(a), the low-energy disorder-induced spin excitations also become highly anisotropic, as shown in Fig. \ref{nematodes}(d). Because of numerical limitations the spin-susceptibility from inhomogeneous real-space configurations is obtained from systems with $N=24$. Surprisingly, the disorder significantly enhances the nematic response as seen by comparing Figs. \ref{nematodes}(e,f); (e) displays two distinct cuts along either $q_x=\pi$ or $q_y=\pi$ from Fig. \ref{nematodes}(d), and reveals a much larger anisotropy in the spin response for the disordered case (e) compared to the clean case (f). Since a gapped spectrum near the Fermi level (caused by superconductivity or pseudo-gap physics) is a prerequisite for disorder-induced magnetization, we obtain the interesting situation where a disordered superconductor in the presence of sub-dominant electronic correlations and a small $x-y$ symmetry breaking field, work as a catalyst for observing a nematic response in neutron scattering.

\begin{figure}[]
\begin{minipage}{.49\columnwidth}
\includegraphics[clip=true,width=0.98\columnwidth]{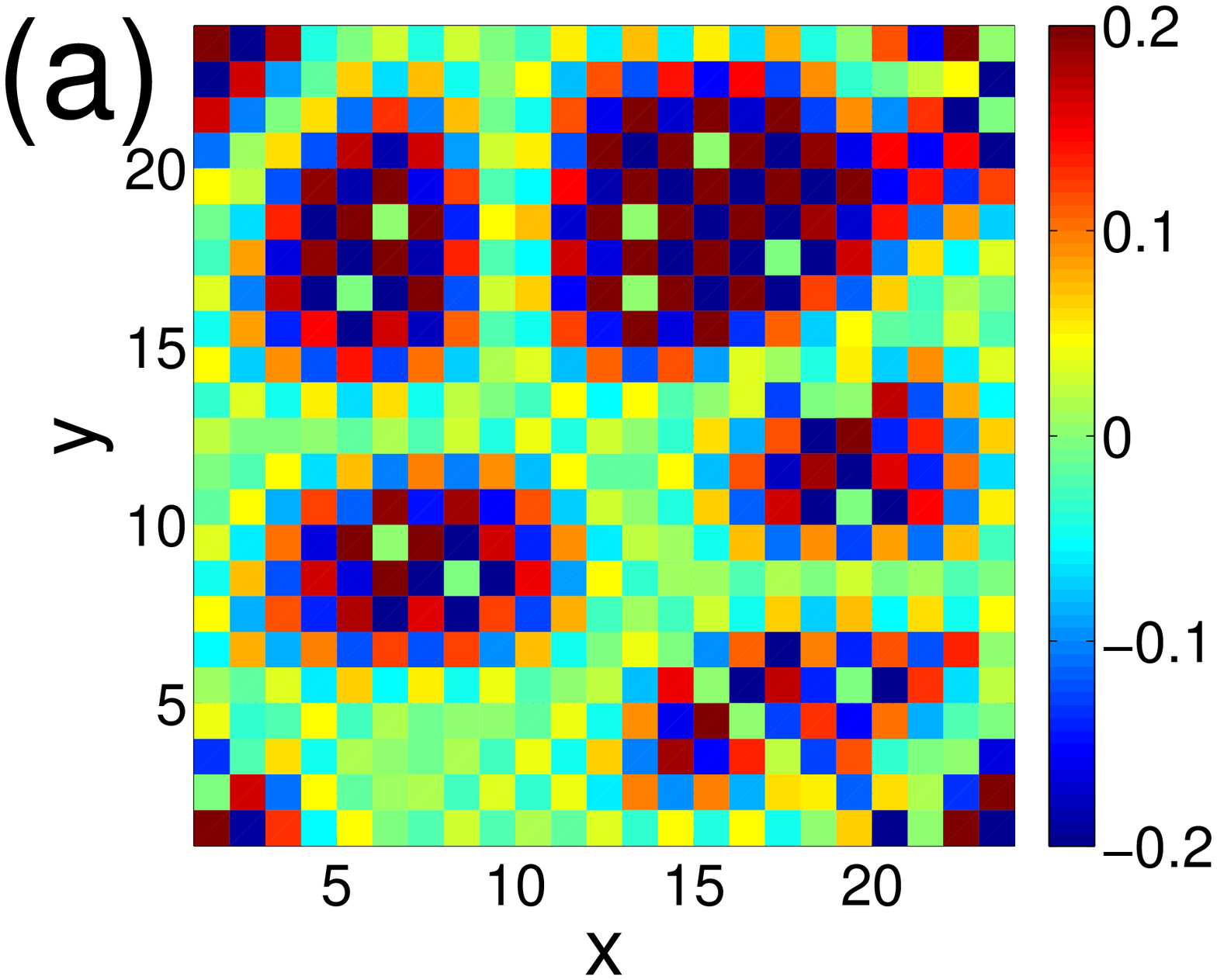}
\end{minipage}
\begin{minipage}{.49\columnwidth}
\includegraphics[clip=true,width=0.98\columnwidth]{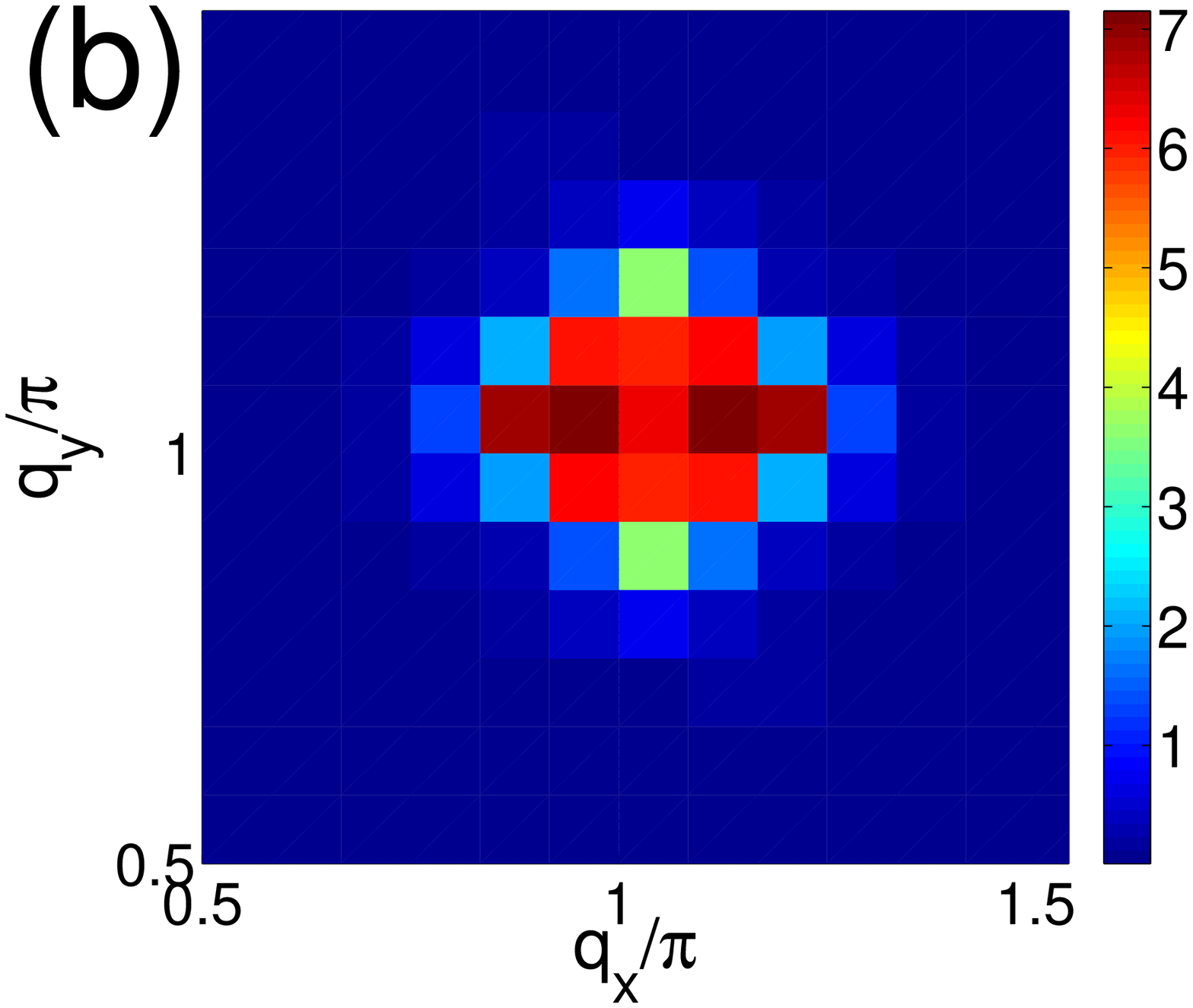}
\end{minipage}
\begin{minipage}{.49\columnwidth}
\includegraphics[clip=true,width=0.98\columnwidth]{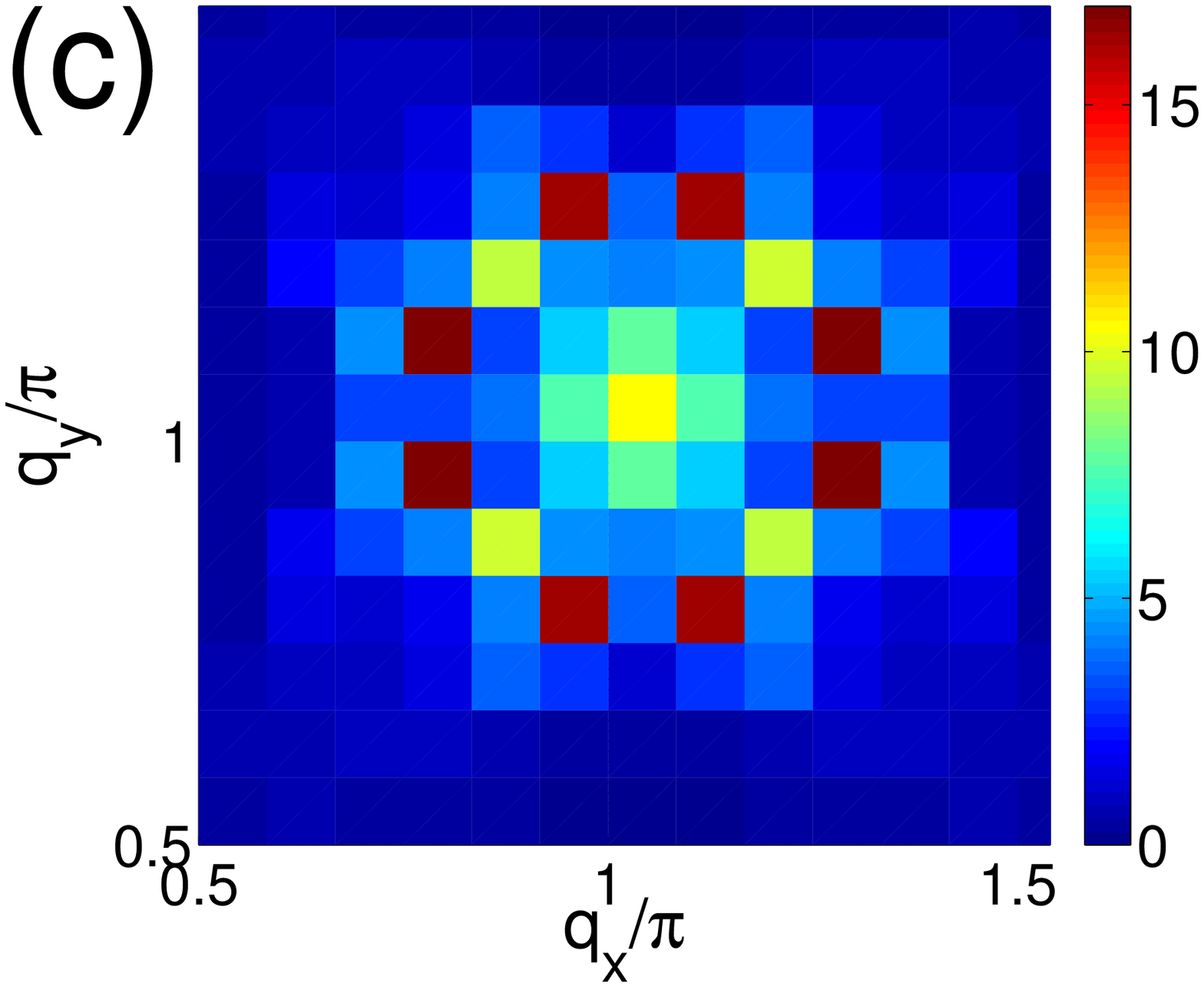}
\end{minipage}
\begin{minipage}{.49\columnwidth}
\includegraphics[clip=true,width=0.98\columnwidth]{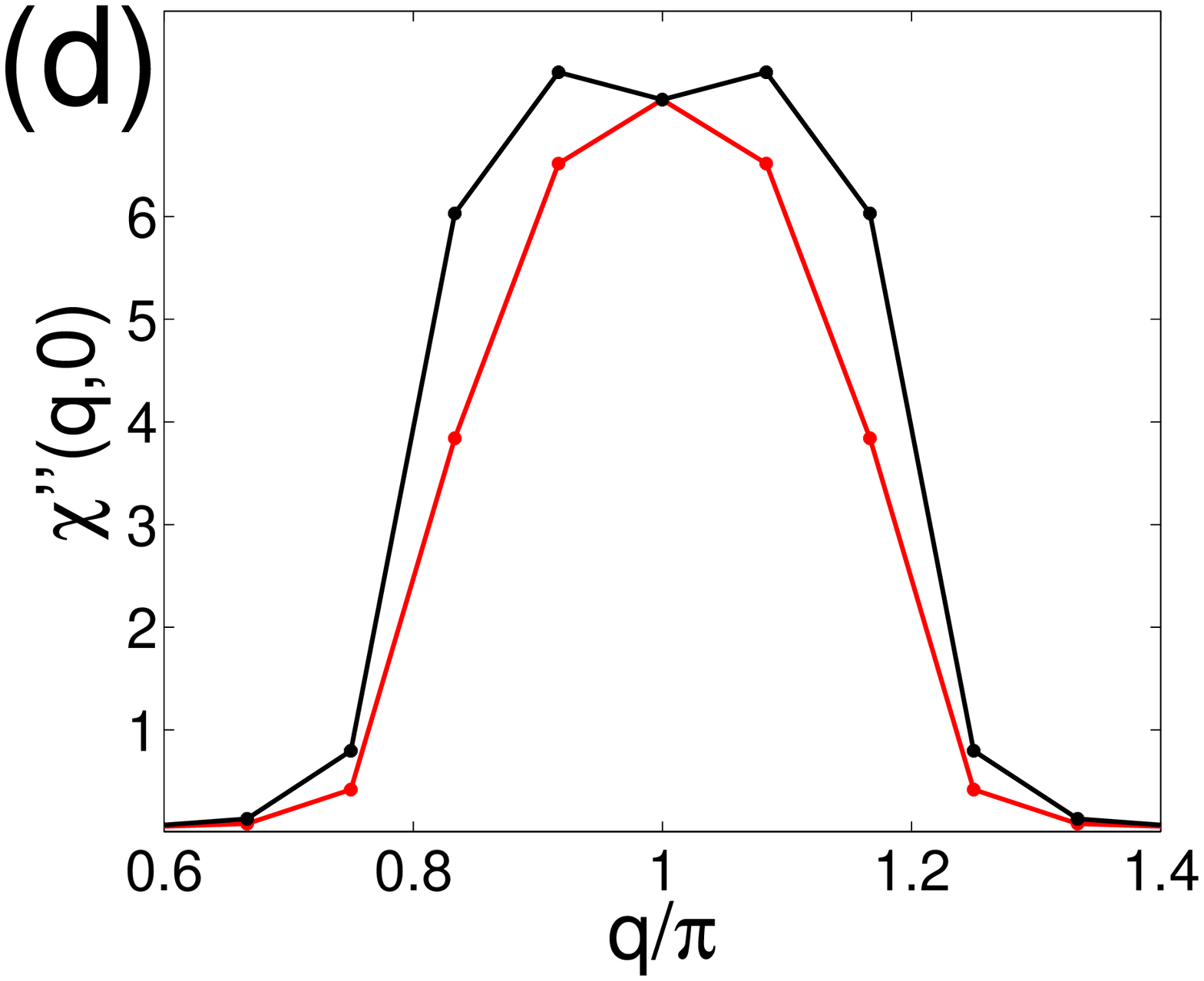}
\end{minipage}
\caption{(Color online) (a) Static magnetization shown in real-space for one of the disordered systems used in (b-d). As seen, the frozen magnetization roughly consists of anti-phase-coupled  antiferromagnetic domains. (b) Constant energy cuts of the susceptibility $\chi''(q,\omega)$ with $U/t=2.5$ and $\omega/t=0.0t$ averaged over ten different $24\times 24$ systems each containing 4\% impurities of strength $V_i=5t$. (c) same as (b) but at $\omega/t=0.40t$. (d) shows $q_x$ ($q_y$) cuts of $\chi''(q,\omega)$ from (b) along $q_y=\pi$ [black line] ($q_x=\pi$ [red line]).  } \label{cutU25asym005disordered}
\end{figure}

We end this section with a brief discussion of the realistic many-impurity situation where the impurity-induced magnetization form a glassy pattern shown in Fig.  \ref{cutU25asym005disordered}(a). Panels (b) and (c) in Fig. \ref{cutU25asym005disordered} displays respectively the low- and high-energy spin susceptibility, $\chi''(q,\omega)$, in the presence of a few percent added disorder similar to the experimental study in e.g. Ref. \cite{suchaneck}. The particular concentration of disorder (and their strength) is not important for the present discussion. In agreement with the neutron measurements, the low-energy spin response from a collection of overlapping spin nematogens peaks at the incommensurate (commensurate) position along $q_x$ ($q_y$). Again this distinct asymmetry is only present in the low-energy sector as seen from Fig. \ref{cutU25asym005disordered}(c). In Fig. \ref{cutU25asym005disordered}(d) we show cuts through Fig. \ref{cutU25asym005disordered}(b) along $q_y=\pi$ [black line] and $q_x=\pi$ [red line], revealing a semi-quantitative fit to the experimental low-energy spin fluctuations obtained in Ref. \cite{hinkov08a}.

\section{Conclusions}
We have calculated the spin susceptibility of a $d$-wave superconductor with Hubbard correlations  in the presence of small explicit symmetry breaking of the underlying lattice. Correlations were found to significantly enhance the low-energy anisotropy of the spin response as the stripe instability is approached. Comparisons with experiment\cite{hinkov08a,suchaneck}, as well as  with work based on  the same model and
 recent analysis of the doping dependence of such effects at strong coupling\cite{rasmus} suggest that this effect is responsible for the strongly enhanced nematic tendency in the spin response observed in the YBCO system as it is underdoped.  In addition, we have shown here that disorder significantly enhances the nematicity, via generation of local spin nematogens which exhibit an anisotropic low-energy spin response that can be significantly enhanced compared to the clean case, and is crucial to understand  the $q$-space form of the neutron response near ($\pi,\pi$)
 observed in experiments on strongly underdoped, untwinned YBCO samples.   Our predictions can be tested by studying the effect of $Zn$ substitution
 on the anisotropic spin response, and by analysis of NMR and STM spectroscopy of impurity bound states in underdoped cuprates.

\acknowledgments
B.M.A. acknowledges support from The Danish Council for Independent Research $|$ Natural Sciences. P.J.H acknowledges support from NSF-DMR-1005625.

\end{document}